\documentclass[prd,twocolumn,nofootinbib,showpacs,floatfix,
superscriptaddress]{revtex4}
 \pdfoutput=1
 
\usepackage{graphicx}
\usepackage{epsfig}
\usepackage{bm}
\usepackage[T1]{fontenc}
\usepackage[latin9]{inputenc}
\usepackage{amssymb}
\usepackage{float}
\usepackage{amsmath}
\usepackage{dcolumn}
\usepackage{cancel}
\usepackage[colorlinks]{hyperref}
\usepackage[usenames,dvipsnames]{color}
\hypersetup{
     breaklinks=true,
    pdfstartview={FitH},    % fits the width of the page to the window
    colorlinks=true,       % false: boxed links; true: colored links
    linkcolor=blue,          % color of internal links
    citecolor=red,        % color of links to bibliography
    filecolor=magenta,      % color of file links
    urlcolor=blue,           % color of external links
    anchorcolor=green,      % Color for anchor text
    linktocpage=true
}

\newcommand{\be}{\begin{equation}}
\newcommand{\ee}{\end{equation}}
\newcommand{\De}{\Delta}

\begin{document}
\title{Baryon asymmetry from Barrow   entropy: theoretical predictions and 
observational constraints}
%%%%%%%%%%%%%%%%%%%%%%%%%%%%%%%%%%%%%%%%%%%%%%%%%%%%%%%%%%%%%%%%%%%%%%%%%%%%%%%%%%%%%%%%%%%%%%%%%%%%%%%%%%%%%%%%%%%

\author{Giuseppe Gaetano Luciano}
\email{gluciano@sa.infn.it}
\affiliation{Dipartimento di Fisica, Universit\`a di Salerno, Via Giovanni Paolo 
II, 132 I-84084 Fisciano (SA), Italy}
\affiliation{ INFN, Sezione di Napoli, Gruppo collegato di Salerno, Via Giovanni 
Paolo II, 132 I-84084 Fisciano (SA), Italy}

\author{Emmanuel N. Saridakis}
\email{msaridak@noa.gr}
\affiliation{National Observatory of Athens, Lofos Nymfon, 11852 Athens, Greece}
\affiliation{Department of Astronomy, School of Physical Sciences,University of 
Science and
Technology of China, Hefei, Anhui 230026, China}
\affiliation{CAS Key Laboratory for Research in Galaxies and 
Cosmology,University of
Science and Technology of China, Hefei, Anhui 230026, China}

\begin{abstract}
 We investigate the generation of baryon asymmetry from   the corrections 
brought about in the Friedman equations due to Barrow entropy. In particular, by 
applying the gravity-thermodynamics conjecture one obtains extra terms in   the 
Friedmann equations that change the Hubble function evolution during the 
radiation-dominated epoch. Hence, even in the case of standard  coupling 
between the Ricci scalar and  baryon current they can lead to a non-zero  
baryon asymmetry. In order to match observations we  find that the Barrow 
exponent should lie in the interval   $0.005\lesssim\De\lesssim0.008$, which 
corresponds to a slight deviation from the standard Bekenstein-Hawking entropy. 
The upper bound is tighter than the one of other observational constraints, 
however the interesting feature is that in the present analysis we obtain a 
non-zero lower bound.  Nevertheless this lower bound would disappear if the 
baryon asymmetry in Barrow-modified cosmology is generated by other mechanisms, 
not related to the   Barrow modification.
\end{abstract}


%\keywords{Barrow entropy, black holes, baryon asymmetry, Friedmann equations} 

\maketitle

\section{Introduction}
\label{Intro}

Since the first investigation of the thermodynamic features
of black holes, the connection between gravity and thermodynamics
has been widely explored, becoming even more central in the recent effort of 
developing a quantum theory of 
gravity~\cite{Bardeen:1973gs,Bekenstein:1973ur,Hawking:1975vcx}. This concept 
was   later formalized 
by the so-called \emph{gravity-thermodynamics} 
conjecture~\cite{Padmanabhan:2003gd,Eling:2006aw,Padmanabhan:2009vy}, 
which states that the field equations of General Relativity may arise from the 
laws of thermodynamics applied on   
 spacetime itself \cite{Jacobson:1995ab}. 
On cosmological grounds, the investigation of the gravity-thermodynamics 
conjecture has revealed that Friedmann equations can be extracted by applying 
the first law of  thermodynamics to the apparent horizon of the 
Universe   
 \cite{Frolov:2002va,Cai:2005ra,Akbar:2006kj,Cai:2006rs}, thus unveiling 
interesting scenarios in a plethora of contexts
\cite{Akbar:2006er,Paranjape:2006ca,Calcagni:2005vn,Sheykhi:2007gi,
Sheykhi:2008qs,CANTATA:2021ktz, Addazi:2021xuf}.

In the commonly adopted formulation,  the gravity-thermodynamics conjecture
makes use of the Bekenstein-Hawking (BH)  area law $S_{BH}=A/(4G)$
for the black-hole horizon entropy ($A=4\pi r_{hor}^2$ is the area with 
$r_{hor}$ the black-hole horizon) and applies it to the Universe apparent 
horizon. However, 
several extensions have appeared in the literature, by using various modified 
entropy relations, arising from    non-extensive generalizations
of the statistics of horizon degrees of freedom
and/or quantum gravitational deformations of the horizon
geometry. Among these, special focus
has been placed on Tsallis~\cite{Tsallis} and Kaniadakis~\cite{Kania1} entropies,
whose cosmological applications have been addressed
in~\cite{Sheykhi:2007zp,Jamil:2009eb,Sheykhi:2018dpn,
SayahianJahromi:2018irq, Saridakis:2018unr, 
Lymperis:2018iuz,Sheykhi:2019bsh,Nunes:2015xsa,Abreu:2021kwu,Lymperis:2021qty, 
Hernandez,Abdalla:2022yfr}.

Recently, a plausible generalized entropy
which is based on  a modified horizon endowed with a 
fractal structure, has been proposed
by Barrow as \cite{Barrow:2020tzx}
\be
\label{BE}
S_\De\,=\,\left(\frac{A}{4G}\right)^{1+\Delta/2}\,.
\ee
In particular, quantum effects are
parameterized by the Barrow exponent $0\le\Delta\le1$, 
with $\Delta=0$ giving the BH
limit, while $\Delta=1$ corresponds to the maximal
deformation. Although  Barrow entropy was 
formulated for black holes, in the lines of gravity-thermodynamic conjecture it 
can be applied in a cosmological framework. In this way, one acquires 
corrections to the Standard Model of Cosmology (SMC), namely on the Friedmann 
equations, brought about by the Barrow entropy  \cite{Saridakis:2020lrg}. 
Additionally, one can apply Barrow entropy to the holographic principle, 
obtaining 
 Barrow  holographic dark
energy 
\cite{Saridakis:2020zol,Dabrowski:2020atl,Saridakis:2020cqq,Sheykhi:2021fwh,
Adhikary:2021xym}. Hence, one can confront the above constructions with 
observational data end amongst others extract constraints on the Barrow exponent 
$\Delta$ 
\cite{Anagnostopoulos:2020ctz,Leon:2021wyx,Barrow:2020kug,Jusufi:2021fek}.
As expected, in all these studies deviations from the BH
entropy are found to be relatively small.
 
Up to now the phenomenology of Barrow-entropy-based cosmological models
mostly deals with  the impositions of constraints on Barrow exponent
by comparison with confirmed  predictions of cosmology. However, it would be 
interesting to examine   whether 
these models can account for observational evidences 
which are not perfectly understood within   standard cosmology. 
In this perspective, one of such   puzzles is the 
origin of Baryon Asymmetry in the Universe (BAU). As it was found by 
Sakharov~\cite{Sakha},   in a particle physics theory  three conditions have 
to be satisfied in order to produce BAU: 
\emph{i)} baryon number $B$ violation,  
\emph{ii)} $C$-symmetry and $CP$-symmetry violation and
\emph{iii)} out-of-thermal-equilibrium interactions. However, 
%and on the assumption
%of a $C$- and $CP$-violating coupling between
%space-time and baryon current, 
the predicted BAU in the SMC is vanishing.
Although several potential explanations have been hitherto
suggested~\cite{Kugo2,Cohen,Basini:2002xb,Davoudiasl, 
Lambiase:2006dq,Lambiase:2006ft,Li:2004hh,Canetti,Alex,Mavromatos:2018map,
Basilakos:2019acj,Arbuzova:2019xti,Azhar:2020coz,Agrawal:2021usq,
Bhattacharjee:2021jwm, Azhar:2021wvx, Sami:2021ufn,Lamb}  (see also models with  
GUT interactions~\cite{Buras1,Georgi,Dimopoulos,Kolb}), none of them offers a 
full and well-accepted solution \cite{Barrow:2022gsu}.

Starting from the above premises, in this work we
analyze baryogenesis in the framework of Barrow cosmology. Since
 \eqref{BE} induces modifications in the Friedmann equations, it   leads 
  to modified energy  density and pressure  
that can comply with all three Sakharov conditions.
As a result, we obtain a non-vanishing ($\Delta$-dependent) 
expression for the baryon asymmetry parameter $\eta$, which allows us to \emph{i)}
account for the origin of BAU and \emph{ii)} constrain Barrow parameter $\Delta$ 
by comparison with current observational bounds on $\eta$.  We emphasize that  
this picture is physically motivated by the fact that 
quantum gravity corrections 
are expected to play a non-trivial role on the apparent horizon in the early 
Universe, with non-negligible effects on the subsequent cosmic evolution too. 
Therefore, while the research on  Barrow-entropy-based cosmology should 
certainly be considered as preliminary, on the other hand it may provide
valuable hints towards understanding the impact of quantum gravity
on horizon properties and related phenomenology. 
 
The present work is organized as follows: in Section \ref{MFE} 
we implement the gravity-thermodynamics conjecture
with Barrow entropy and  we derive the modified Friedmann equations.
 Then in Section \ref{CTC} we provide a detailed investigation of 
the baryogenesis procedure in Barrow cosmology, and we extract the constraints 
on Barrow exponent $\Delta$.
Conclusions and outlook are finally discussed in Section \ref{DC}.
Throughout the manuscript, we work in natural units $\hbar=c=k_B=1$, while we 
keep the gravitational constant $G$ explicitly. 
 
\section{Friedmann equations in Barrow Cosmology}
\label{MFE}

In this section we present a derivation of Friedmann equations for the 
Friedmann-Robertson-Walker (FRW) metric within the framework of Barrow 
cosmology. Towards this end, we consider the $(1+3)$-dimensional line element
\be
ds^2\,=\,h_{bc} dx^{b}dx^{c}+\tilde r^2(d\theta^2+\sin^2\theta d\phi^2)\,,
\ee
where $h_{bc}=\mathrm{diag}\left[-1,a^2/(1-kr^2)\right]$ $(b,c=\{0,1\})$  is the 
metric
of a $(1+1)$-dimensional subspace of coordinates $x^b\equiv(t,r)$
and  $\tilde r=a(t)r$, with $a(t)$ being the time-dependent scale factor.  Here, 
we have denoted by $k$ the (constant) spatial curvature. 
Moreover, we assume that the Universe is bounded by the apparent
horizon of radius $\tilde r_H=1/\sqrt{H^2+k/{a^2}}$, where 
$H=\dot a(t)/a(t)$ is the Hubble parameter (in the following dots denote 
time-derivatives). 
From the definition of surface
gravity $\kappa$ on the apparent horizon, 
it is easy to show that the related temperature
is~\cite{Akbar:2006kj}
\be
\label{T}
T\,=\,\frac{\kappa}{2\pi}=-\frac{1}{2\pi \tilde r_H}\left(1-\frac{\dot{\tilde 
r}_H}{2H\tilde r_H}\right). 
\ee

We describe the   content 
of the Universe as a perfect fluid of 
energy-momentum tensor
\be
T_{\mu\nu}\,=\,(\rho+p)u_\mu u_{\nu}+pg_{\mu\nu}\,,
\ee 
where $\rho$ and $p$ are the energy density
and pressure at equilibrium, while $u_{\mu}$ is the four-velocity of the fluid. 
The conservation equation
$\nabla_{\mu}T^{\mu\nu}=0$ for the FRW geometry then implies
the continuity equation $\dot \rho=-3H(\rho+p)$. We mention that this fluid can 
be the total (matter plus radiation) one, the matter one or the radiation one, 
according to the application we are interested in each time.

Using the gravity-thermodynamics conjecture, the
cosmological (Friedmann) equations can be derived
by considering the Universe as a thermodynamic system bounded by the apparent 
horizon
and applying the first law of thermodynamics 
\be
\label{FLT}
dE\,=\,TdS+WdV
\ee
on the horizon. Here, $E=\rho V$ 
is the total energy of the Universe of
$3$-dimensional volume $V=4\pi \tilde r_H^3/3$
and surface area $A=4\pi\tilde r_H^2$, while $S$ denotes
the horizon entropy. 
Since the Universe is
expanding, a work   $W=-\frac{1}{2}T^{bc}h_{bc}=\frac{1}{2}(\rho-p)$
is associated to the change of volume $dV$. Concerning the entropy $S$, in 
general it is an extension of the  Bekenstein-Hawking entropy, and focusing on 
modifying area laws it can be written as 
\be
\label{Sbis}
S\,=\,\frac{f(A)}{4G},
\ee
where the   function $f(A)$ quantifies the deviation from the standard BH 
relation (the latter is recovered for $f(A)=A$).
Insertion of the above relations into  \eqref{FLT} leads to 
the first Friedmann equation \cite{Saridakis:2020lrg}
\be
\label{fFE}
-4\pi G \left(\rho+p\right)\,=\,\left(\dot H-\frac{k}{a^2}\right)f'(A)\,,
\ee
where the prime denotes    derivative 
with respect to $A$. Moreover, by utilizing the continuity equation and 
integrating we obtain the second Friedmann equation as
\be
\label{SecEq}
\frac{8\pi G}{3}\rho\,=\,-4\pi\int \frac{f'(A)}{A^2}dA\,.
\ee
Note that    Eqs.~\eqref{fFE} and~\eqref{SecEq} 
are of general validity, and the specif features  of the adopted entropic
model are quantified by the function $f(A)$    in  \eqref{Sbis}.
As expected in the  case  $f(A)=A$ of BH entropy, we recover the standard 
  Friedmann equations.
  
  In the case of  Barrow entropy \eqref{BE} we obtain 
the  modified Friedmann equations \cite{Saridakis:2020lrg}
\begin{eqnarray}
\label{FME}
-4\pi 
G\left(\rho+p\right)&\hspace{-0.5mm}=\hspace{-0.5mm}&\alpha_\De\!\left(\dot 
H-\frac{k}{a^2}\right)\!\left[G\!\left(H^2+\frac{k}{a^2}\right)\!\right]^{-\De/2
},
\\[2mm]
\nonumber
\frac{8\pi 
G}{3}\rho&\hspace{-0.5mm}=\hspace{-0.5mm}&\frac{2\alpha_\De}{G\left(2-\De\right)
}\left[G\left(H^2+\frac{k}{a^2}\right)\right]^{1-\De/2}\,+\,c\,, \\
\label{SME}
\end{eqnarray}
with 
\be 
\alpha_\Delta\,=\,\frac{\pi^{\De/2}(2+\De)}{2}\,.
\ee   Finally, note that 
 the integration constant $c$ is identified with  the 
cosmological through  
$c=8\pi G\Lambda/{3}$. As mentioned above, in the limit $\De\rightarrow0$, in 
which Barrow entropy recovers BH entropy, the above expressions reproduce the 
standard Friedmann equations.
  
In the following for simplicity we will focus on the flat case   $k=0$, and 
since we are interested in the early universe we consider that the Universe 
cosmic fluid corresponds to the radiation sector and we neglect the 
cosmological constant.

\section{Baryon Asymmetry from Barrow entropy}
\label{CTC}

The origin of matter-antimatter asymmetry in the early Universe
is one of the most debated problems in present day cosmology. 
Observations unambiguously indicate that the
amount of matter prevails over antimatter, 
in constrast to the predictions of the Standard Model of Particle Physics  
  \cite{Canetti}. As we mentioned in the Introduction,
  in order  to generate
 dynamically  the Baryon Asymmetry in the Universe (BAU), the 
  three Sakharov conditions should be fulfilled.
  
The first Sakharov criterion ensures that the Universe
evolves from an originally baryon-symmetric 
state into a configuration where the difference
\be
\label{quant}
\frac{\eta}{7}\,\equiv\,\frac{n_B-n_{\bar{B}}}{s}%\,=\, \left|
%\frac{15\,g_b}{4\pi^2\,g_{*s}}\frac{\mathcal{\dot R}}{M_*^2\, T}\right|\,,
\ee
is no longer vanishing, with  $n_B$ ($n_{\bar{B}}$) being 
the baryon (anti-baryon) number density, and $s$
  the entropy density in the radiation-dominated era. 
The second Sakharov condition
is required due to the fact that  if $C$ and $CP$ were exact Hamiltonian 
symmetries
then the total rate for any interaction producing 
an excess of baryons would be compensated by the 
complementary process producing an excess of anti-baryons.
Finally, the last condition can be understood by calculating the equilibrium
average of baryon number $B$ as \cite{Trodden:2004mj}
\begin{eqnarray}
\nonumber
\langle B\rangle_\beta&=&\mathrm{Tr}\left(e^{-\beta H} B\right)\,=\,
\mathrm{Tr}\left[\left(CPT\right)\left(CPT\right)^{-1}e^{-\beta H} 
B\right]\\[2mm]
\nonumber
&=& \mathrm{Tr}\left[e^{-\beta H} 
\left(CPT\right)^{-1}B\left(CPT\right)\right]=- \mathrm{Tr}\left(e^{-\beta 
H}B\right)\,,\\
\end{eqnarray}
where $T$ is the time-reversal, and where we have exploited
the fact that $H$ commutes with $CPT$. As a result, one obtains
$\langle B\rangle_\beta=0$ at equilibrium, thus preventing net baryon number 
generation. 
While complying with the first two Sakharov conditions, 
standard cosmology fails to predict baryogenesis,
since the last criterion is not  satisfied during the whole radiation-dominated 
era.

One can satisfy the first  Sakharov condition   within certain supergravity 
theories, as highlighted in~\cite{Kugo2}. 
%On the other hand, the possibility to have BAU
%while maintaining thermality has been discussed in. 
In this framework the $CP$-violating
interaction in vacuum between the derivative of the Ricci scalar
curvature $\mathcal{R}$ and the baryon number current $J^{\mu}$
takes the form~\cite{Davoudiasl,Feng:2004mq}
\be
\label{Jmu}
\frac{1}{M_*^2}\int d^4x\sqrt{-g}\,J^{\mu}\partial_{\mu}\mathcal{R}\,,
\ee
where $M_*= (8\pi G)^{-1/2}$ %\simeq2.4\times10^{18}\,\mathrm{GeV}$ 
is the characteristic cutoff scale (see 
also~\cite{Bento:2005xk,Sadjadi:2007dx,Davoudiasl:2013pda,Oikonomou:2016jjh,
Odintsov:2016hgc,Datta:2020bht}). 
In order to create asymmetry one assumes that there exists some interaction 
violating the baryon number $B$. By noticing that the 
spatial part of $\mathcal{R}$ vanishes for the FRW metric, one has
\be
\label{oeq}
\frac{1}{M_*^2}J^{\mu}\partial_{\mu}\mathcal{R}\,=\,\frac{1}{M_*^2}\left(n_B-n_{
\bar{B}}\right)\mathcal {\dot R}\,.
\ee
Thus, in an expanding
Universe where $\mathcal{R} $ and $\mathcal{\dot R}$ are non-zero, the 
interaction \eqref{Jmu} can produce opposite
energy contributions that differ for particles and antiparticles,  i.e the 
above   gravitational baryogenesis  can generate the baryon-anti-baryon 
asymmetry. Hence, in 
this way one 
obtains a dynamical violation of $CPT$ symmetry,
which affects thermal equilibrium distributions through an effective 
chemical potential $\mu_B=-\mu_{\bar B}=-{\mathcal{\dot 
R}}/{M_*^2}$~\cite{Davoudiasl}.

Once the temperature drops below the decoupling value $T_D$, 
the Universe is driven towards a
non-zero equilibrium asymmetry 
\be
\label{asym}
n_B-n_{\bar{B}}\,=\,\bigg|\frac{g_b}{6}\mu_B T^2\bigg|\,,
\ee
where $g_b\sim \mathcal{O}(1)$ is the
number of the intrinsic degrees of freedom of baryons. 
Using  \eqref{quant}, the 
baryon asymmetry in the standard notation then
reads
\be
\label{quantbis}
\frac{\eta}{7}\,=\,\bigg|\frac{15\,g_b}{4\pi^2\,g_{*s}}\frac{\mathcal{\dot 
R}}{M_*^2\, T}\bigg|_{T=T_D}\,,
\ee
where we have used
\be 
s=\frac{2\pi^2g_{*s}T^3}{45}
\ee
for the entropy density, and with $g_{*s}$   the 
number of degrees of freedom for particles contributing to the entropy of the 
Universe in the radiation-dominated era. 
Note that this number is roughly equal to the total
number $g_*\simeq106$ of degrees of freedom of relativistic Standard Model
particles, as discussed in~\cite{Kolb}.

The essence of expression \eqref{quantbis} is that 
baryon asymmetry can indeed occur as long as the Ricci scalar ${\mathcal{R}}$ 
varies over time.
However, since ${\mathcal{R}}=12H^2+6\dot{H}$,   in the case of standard 
cosmology during the radiation-dominated era, i.e. where $p=\rho/3$, the 
standard Friedmann equations give  ${\mathcal{\dot R}}=0$, and thus   baryon 
asymmetry cannot arise. Nevertheless, as we will show in the following, this 
is not the case in Barrow cosmology, and baryon asymmetry through  
\eqref{quantbis} can indeed arise.

 Let us denote by ${\bar{\rho}}=3H^2/(8\pi G)$ and  ${\bar{p}}={\bar{\rho}}/3$ 
the standard-cosmology energy density and pressure during the radiation era  and 
by $\delta\rho_\De$, $\delta p_\De$ the corresponding Barrow-entropy-induced 
extra terms in the Friedmann equations \eqref{FME} and~\eqref{SME}. Thus, we 
can write  
\begin{eqnarray}
\label{drho}
\delta\rho_\De&=&\left[-1+\beta_\De\left(G^2{\bar{\rho}}\right)^{-\De/2}
\right]{\bar{\rho}}\,,
\\[2mm]
\label{dp}
\delta p_\De&=&\left[-1+\gamma_\De\left(G^2{\bar{\rho}}\right)^{-\De/2}
\right]\frac{{\bar{\rho}}}{3}\,,
\end{eqnarray}
where 
\be
\beta_\De\,=\,\frac{3^{\De/2}\left(2+\De\right)}{2^{3\De/2}\,\left(2-\De\right)}
\,,\,\,\quad \gamma_\De\,=\,\beta_\De\left(1-2\De\right).
\ee
%Here, we have used the standard 
%form of the second Friedmann equation, 
%since corrections to the latter would only contribute
%to orders higher than that considered in the present analysis (see below).
%From Eqs.~\eqref{drho} and~\eqref{dp}, it follows that Barrow entropy 
%could indeed be responsible for the breakdown
%of thermal equilibrium, thus satisfying the third and last Sakharov condition. 
As expected,  for 
$\Delta\rightarrow0$ we acquire  $\delta\rho_\De, \delta p_\De\rightarrow0$, 
consistently with the 
recovery of   standard cosmology in this limit.  

From the above we can clearly see that in the scenario at hand the Ricci scalar 
is non-zero during the radiation-dominated epoch, and in particular its 
time-derivative is given by 
\be
\label{dotR}
\mathcal{\dot{R}}_\De =\left[\frac{\pi^{3/2}\, 2^{\left(13-3\De\right)/2} 
\De\left(2+\De\right)}{3^{\left(1-\De\right)/2}} \right]
{G^{3/2-\De}} \,{{\bar{\rho}}^{\left(3-\De\right)/2}}\,,
\ee
where  we have implemented the continuity equation at equilibrium. 
Substitution of \eqref{dotR} into~\eqref{quantbis} then gives
\begin{eqnarray}
\label{etafin}
&&
\!\!\!\!\!\!\!\!\!\!\!\!\!\!\!\!\!\!
\eta_\De\,=\left[ \frac{35\, 2^{3\left(3-\De\right)/2}\, 3^{
\left(1+\De\right)/2}\, \De\left(2+\De\right)}{\pi^{1/2}}
\right]\nonumber\\
&&\cdot
 \frac{g_b}{g_* M_*^2\,T_D}\, {G^{
3/2-\De}} {{\bar{\rho}}^{\left(3-\De\right)/2}}\,,
\end{eqnarray}
where ${\bar{\rho}}$ must be calculated at the decoupling point.
 This expression can be   simplified
by expressing the gravitational constant in terms of the Planck mass, 
$G=1/M_P^2$ in our units, and replacing the equilibrium density
by ${\bar{\rho}}|_{_{T=T_D}}=\pi^2 g_* T_D^4/30$, obtaining
\be
\label{etafinbis}
\eta_\De\,=\,\xi_\De\,g_b\hspace{0.7mm}g_*^{\left(1-\De\right)/2}
\left(\frac{T_D}{M_P}\right)^{5-2\De}\,,
\ee
where
\be
\xi_\De\,=\,\frac{7\,\pi^{7/2-\De}\,2^{6-\De}\,\De\left(2+\De\right)
}{3^{1-\De}\times5^{\left(1-\De\right)/2}}\,.
\ee
Finally, since the Barrow exponent has been found by various studies to satisfy 
 $\De\ll1$ 
\cite{Anagnostopoulos:2020ctz,Leon:2021wyx,Barrow:2020kug,Jusufi:2021fek}, 
which is expected since Barrow entropy should not deviate significantly from 
Bekenstein-Hawking one,  we can expand the above expression resulting to 
\be
\label{leado}
\eta_\De\,=\,\frac{896\hspace{0.4mm}\pi^{7/2}\hspace{0.4mm}\De}{3\sqrt{5}}\,{
g_b\,\sqrt{g_*}}\,\left(\frac{T_D}{M_P}\right)^5\,+\,\mathcal{O}(\De^2)\,.
\ee
Hence, one can clearly see that the Barrow exponent can lead to a non-zero 
baryon asymmetry, due to te corrections in the Friedmann equations.

\begin{figure}[ht]
    \centering
 \hspace{-0.5cm}   \includegraphics[scale=0.235]{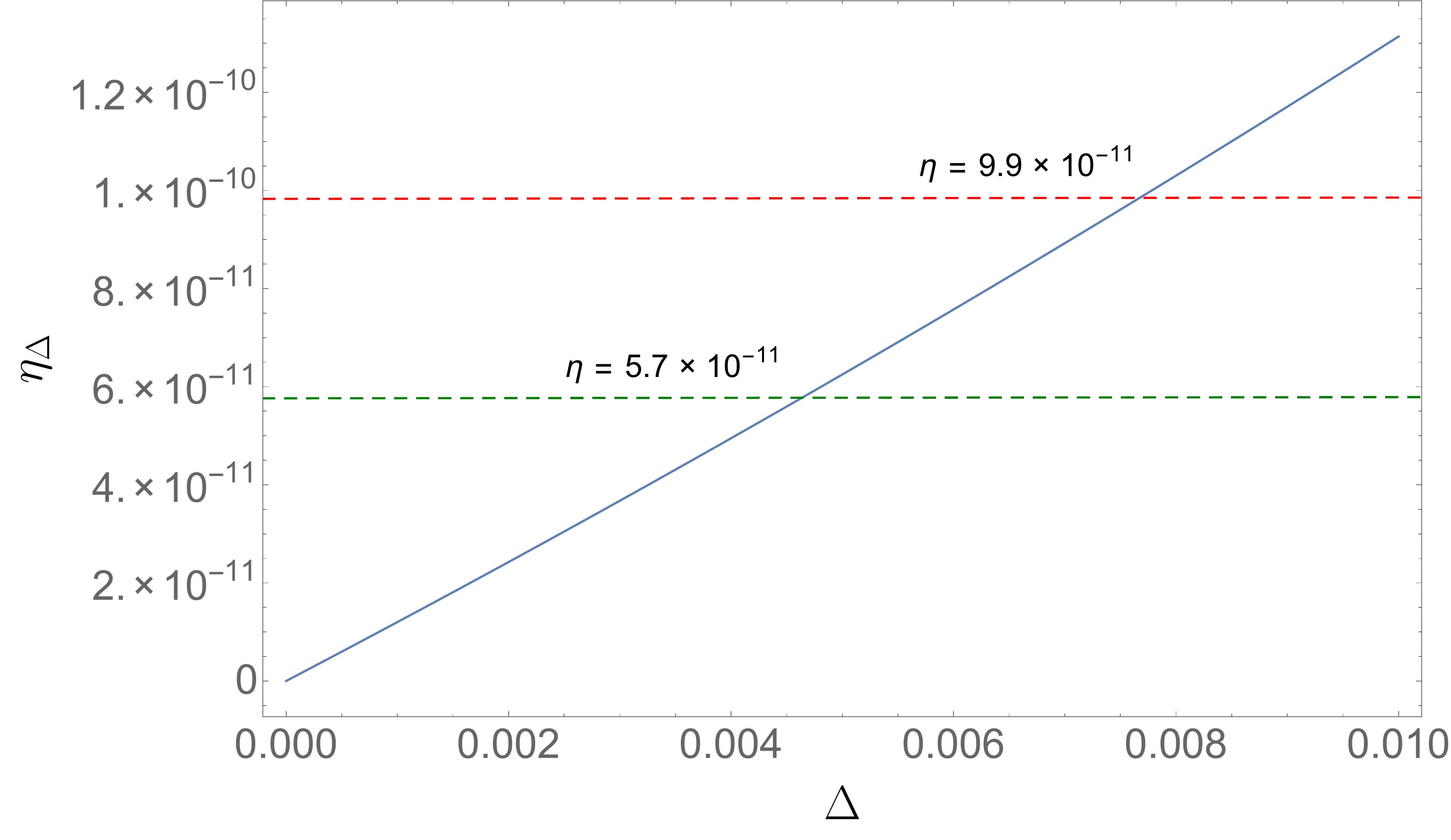}
    \caption{{\it{ The baryon asymmetry    $\eta_\De$ (blue solid 
curve) in the 
case of Barrow-entropy-based cosmology, as a function of the Barrow exponent 
$\Delta$, according to the exact expression \eqref{etafinbis}. 
The
red and green dashed lines mark the observational
upper and lower bounds on $\eta$, respectively.    }}}
    \label{plot}
\end{figure}

In order to proceed to quantitative calculations we consider as usual the 
decoupling temperature  to be $T_D\simeq M_I$, 
where $M_I\sim3.3\times10^{16}\,\mathrm{GeV}$ is the upper
bound on tensor mode fluctuations   at inflationary scale \cite{Lamb}. 
In Fig.~\ref{plot} we depict the prediction for the baryon asymmetry in the 
case of Barrow-entropy-based cosmology, as a function of the Barrow exponent 
$\Delta$, according to the exact expression \eqref{etafinbis}. Additionally, in 
the same figure we also  present the observational bounds on $\eta$ arising 
from  baryogenesis, namely 
\cite{Lamb,Estimate1,Estimate2,Estimate3,Estimate4,Estimate5}: 
\be
\label{ulb}
5.7\times10^{-11}\lesssim\eta\lesssim9.9\times10^{-11}\,.
\ee 

As we can observe, if we desire the baryon asymmetry to originate from the 
effects of the Barrow entropy in the Friedmann equations, we must require
  $0.005\lesssim\De\lesssim0.008$. This interval is tighter than the one 
arising 
from    cosmological datasets from Supernovae (SNIa)
Pantheon sample and cosmic chronometers, namely 
$\Delta=0.0094^{+0.094}_{-0.101}$ \cite{Anagnostopoulos:2020ctz,Leon:2021wyx}, 
as well as from the one obtained from  M87* and S2 star observations, namely
  $\Delta=0.0036^{+0.0792}_{-0.0145}$ \cite{Jusufi:2021fek}, and it is slightly 
wider than the one from   Big Bang Nucleosynthesis (BBN), i.e. 
$\Delta\lesssim10^{-4}$~\cite{Barrow:2020kug}.
However, apart from obtaining a tight upper bound, the important feature of the 
present analysis, contrary to the
other datasets, is that we obtain a non-zero lower bound. Definitely this lower 
bound would disappear if we do not require the baryon asymmetry to arise due to 
the extra terms in the Friedman equations, even if we do have Barrow-modified 
cosmological equations (i.e. one could have the case of Barrow-modified 
cosmology in which  baryon asymmetry is generated by other mechanisms that have 
been proposed in the literature, not related to  Barrow-modified cosmology 
itself).

\section{Discussion and Conclusions}
\label{DC}

Observations reveal a baryon asymmetry that cannot be easily explained  in 
the framework of standard cosmology, and thus offer  an indication that some 
form of   new physics might be needed. In this work we 
investigated the   generation of baryon asymmetry due to 
the corrections brought about in the Friedman equations due to Barrow entropy.

Barrow entropy is  a modified entropy arising from
quantum-gravitational effects on the Universe horizon, 
quantified by the new parameter $0\le\Delta\le1$. By applying the 
gravity-thermodynamics conjecture one obtains extra terms in   the Friedmann 
equations. These extra terms change the Hubble function evolution during the 
radiation-dominated epoch, and  hence even in the case of 
the standard  coupling between the Ricci scalar and  baryon current they can 
lead to a non-zero  baryon asymmetry.  We mention that a similar analysis of 
baryogenesis motivated by quantum gravity phenomenology
has been carried out in the context of deformed uncertainty 
relations \cite{Lamb}.

As we showed, if we desire the baryon asymmetry to be generated from the 
Barrow-entropy-based modified cosmology then we should have a Barrow exponent 
bounded in the interval   $0.005\lesssim\De\lesssim0.008$, which corresponds to 
a slight deviation from the standard Bekenstein-Hawking entropy. The 
upper bound is tighter than the one of other observational constraints, 
however the interesting feature is that in the present analysis we obtain a 
non-zero lower bound.  Nevertheless this lower 
bound would disappear if the baryon asymmetry in Barrow-modified 
cosmology is generated by other 
mechanisms, not related to the   Barrow modification.

Finally, following the analysis of 
\cite{Nojiri:2019skr,Luciano:2021mto,Luciano:2021onl,Jizba} for
Tsallis thermodynamics, it would be interesting
to investigate the extended model where the Barrow exponent has a running (i.e. 
time-dependent) behavior. Although not contemplated in the original formulation 
by Barrow,  this feature might provide an interesting  explanation on why 
datasets from different cosmological eras provide different bounds on $\De$. 
Such investigations lie beyond the scope of the present work, and are left for 
future projects.

\begin{acknowledgments} 
The authors 
acknowledge participation in the COST Association Action CA18108 ``Quantum 
Gravity Phenomenology in the Multimessenger Approach''.
\end{acknowledgments}

\end{document}